\documentclass[aip,jcp,amsmath,amssymb,reprint,superscriptaddress]{revtex4-1}

\usepackage{nameref}
\usepackage{titlesec}
\usepackage{graphicx}
\usepackage{xcolor}
\usepackage{dcolumn}
\usepackage{bm}
\usepackage[utf8]{inputenc}
\usepackage[T1]{fontenc}
\usepackage{mathptmx}
\usepackage{etoolbox}
\usepackage{times}
\usepackage[normalem]{ulem}
\usepackage[colorlinks=true]{hyperref}
\hypersetup{
    colorlinks=true,
    linkcolor=blue,
    filecolor=magenta,     
    citecolor=brown,
    urlcolor=cyan,
    pdftitle={Overleaf Example},
    pdfpagemode=FullScreen,
}

\makeatletter
\def\@email#1#2{%
 \endgroup
 \patchcmd{\titleblock@produce}
  {\frontmatter@RRAPformat}
  {\frontmatter@RRAPformat{\produce@RRAP{*#1\href{mailto:#2}{#2}}}\frontmatter@RRAPformat}
  {}{}
}%
\makeatother

\newcommand{\kT}{k_{\mathrm{B}}T}

\newcommand{\dd}{\mathrm{d}}
\newcommand{\zmax}{z_{\max}}

\newcommand{\dF}{\Delta F^{\!*}}

\newcommand*{\figref}[1]{%
  \hyperref[{#1}]{%
    Figure~\ref*{#1}%
  }%
}
\renewcommand*{\eqref}[1]{%
  \hyperref[{#1}]{%
    eq~\ref*{#1}%
  }%
}
\newcommand*{\tabref}[1]{%
  \hyperref[{#1}]{%
    Table~\ref*{#1}%
  }%
}

\begin{document}

\title[Self-adaptive tube restraint for path collective variables]{A self-adaptive
tube restraint for free-energy calculations along path collective variables}

\author{Radu A. Talmazan}
\affiliation{Laboratoire de Physique et Chimie Th\'eoriques, Unit\'e Mixte de
Recherche n$^{\circ}$7019, Universit\'e de Lorraine, B.P. 70239,
54506 Vand\oe uvre-l\`es-Nancy cedex, France}
\author{Cheng Giuseppe Chen}
\affiliation{Laboratoire de Physique et Chimie Th\'eoriques, Unit\'e Mixte de
Recherche n$^{\circ}$7019, Universit\'e de Lorraine, B.P. 70239,
54506 Vand\oe uvre-l\`es-Nancy cedex, France}
\author{Chenyu Tang}
\affiliation{Laboratoire de Physique et Chimie Th\'eoriques, Unit\'e Mixte de
Recherche n$^{\circ}$7019, Universit\'e de Lorraine, B.P. 70239,
54506 Vand\oe uvre-l\`es-Nancy cedex, France}
\author{Christophe Chipot}
\email{chipot@illinois.edu}
\affiliation{Laboratoire de Physique et Chimie Th\'eoriques, Unit\'e Mixte de
Recherche n$^{\circ}$7019, Universit\'e de Lorraine, B.P. 70239,
54506 Vand\oe uvre-l\`es-Nancy cedex, France}
\affiliation{Department of Biochemistry and Molecular Biology, University of
Chicago, Chicago, Illinois 60637, USA}
\affiliation{Department of Chemistry, University of Chicago, Chicago,
Illinois 60637, USA}
\affiliation{Theoretical and Computational Biophysics Group, Beckman Institute,
and Department of Physics, University of Illinois at Urbana-Champaign, Urbana,
Illinois 61801, USA}

\date{\today}

\begin{abstract}
Path collective variables (PCVs) reduce a high-dimensional transition to a
progress coordinate, $s$, and an orthogonal distance, $z$. Computing the free energy along $s$ often requires restraining $z$, so that sampling embraces a tube centered on the reference path. However,
a conventional fixed half-harmonic wall demands an a~priori tube width
that is both system dependent and $s$-dependent. Too tight a tube biases the
path-projected free energy. Conversely, too loose a tube induces numerical instability and inter-channel leakage. We present an adaptive tube restraint the half-width of which evolves on the fly to track a fixed contour $\dF$ of the orthogonal free energy
$F(z\mid s)$, widening automatically in flat basins and narrowing at pinched
saddles, with minimalist user intervention. We prove that, for a locally harmonic perpendicular well, a contour-following tube captures an $s$-independent fraction of the orthogonal partition function, so that in the hard-wall limit, its bias cancels out. In stark contrast, the bias of a fixed-width tube varies with $s$ and distorts the free-energy landscape. We probe the algorithm on several model potentials, on
$N$--Acetyl--$N^\prime$--methylalanylamide isomerization, on the folding of the mini-protein chignolin, and on the folding-upon-binding of the ribose-binding protein. The method is implemented in the open-source Colvars library, and is, therefore, usable within popular MD engines, such as NAMD, LAMMPS, GROMACS, and Tinker-HP.
\end{abstract}

\maketitle

\section{Introduction \label{Sec:Intro}}


Much of what we want to know about a molecular system is encoded in free-energy
differences between metastable states, from the relative stability of two
conformations to the height of the barrier separating them. The transitions between these
states are rare on the timescale of unbiased molecular dynamics, so the
configurations that matter are sampled too infrequently to be characterized directly.
Enhanced-sampling methods address this shortcoming by adding a bias along a small set of
collective variables (CVs) chosen to capture the slowest relevant motions, while recovering the free
energy as a function of these variables.\cite{chipot2014,henin2022} Chief among these methods are metadynamics and its 
variants\cite{laio2002,laio2008,barducci2008,valsson2016,valsson2014,invernizzi2020}
and the adaptive biasing force (ABF) method with its extended-Lagrangian
ansatz.\cite{darve2001,darve2008,comer2015,lesage2017,fu2016,fu2018} However, the
result is only as good as the underlying CV.\cite{besthummer2005,fiorin2013}
A well-chosen CV resolves the transition, whereas a poor one leaves slow
degrees of freedom unbiased and the free energy poorly converged. 

A natural
choice for a transition between two known metastable states is a coordinate
that follows a reference path.
This path can be defined as an ordered set of configurations approximating the route
connecting the two metastable states. Under these premises, a path collective variable (PCV)
reports two numbers: (i) A progress coordinate, $s$, measures the position along the path, and (ii) an
orthogonal distance, $z$, measures the perpendicular excursion from it. PCVs are widely used in computational studies, from conformational transitions\cite{jiang2022} and ligand
reversible binding\cite{saladino2012} to chemical reactions.\cite{rivasfernandez2026,Talmazan2026} The
free-energy landscape of interest is encapsulated in the profile $F(s)$, which is well defined only when sampling populates the region of the
path.

This requirement creates a central practical difficulty in PCV-based enhanced sampling. If the simulation explores configurations far from the path, the mapping to $s$ becomes geometrically ill-conditioned. Without a nearby reference image providing a reliable local frame, the PCV gradients can become unstable, and the progress coordinate can lose the physical meaning. Conversely, if multiple transition channels share the same range of $s$, but differ in their orthogonal displacement $z$, biasing $s$ alone can encourage the trajectory to switch between channels. The recovered $F(s)$ then averages over mechanisms that should be treated separately. Therefore, restraints on $z$ are needed, not merely for numerical stability, but also for defining the thermodynamic ensemble being sampled.

Orthogonal confinement is therefore common in string-method calculations\cite{weinan2002,estring2005,maragliano2006} and in PCV sampling, often through a root-mean-square-deviation (RMSD) restraint or a fixed half-harmonic wall at a prescribed distance $\zmax$. Yet this fixed-width choice is intrinsically unsatisfactory. If the tube is set to be too narrow, it truncates natural perpendicular fluctuations and introduces an $s$-dependent entropic bias into $F(s)$. If the tube is too wide, it fails to prevent geometric instability or leakage between neighboring channels. The appropriate width is generally not constant as basins may tolerate broad orthogonal fluctuations, whereas saddle regions or narrow passages may require tighter confinement. Because this geometry is usually unknown before sampling, choosing $\zmax$ becomes a system-dependent tuning problem.

Here in this work, we introduce an adaptive confinement tube, the half-width of which is determined locally along the path. At each value of $s$, the wall position is adjusted to follow a prescribed free-energy excess $\dF$ of the conditional orthogonal free energy $F(z\mid s)$. As shown in \nameref{Sec:Theory}, for a locally harmonic orthogonal well, a tube defined by a fixed free-energy contour captures an $s$-independent fraction of the orthogonal partition function. In the hard-wall limit, confinement therefore contributes only an additive constant to $F(s)$, leaving free-energy differences along the path unchanged. The method replaces an ambiguous geometric cutoff in $z$, with a physically interpretable free-energy scale, $\dF$, that defines the allowed orthogonal fluctuations along the path.

The present approach belongs to a broader class of methods that regulate the accessible region of configurational space. Funnel metadynamics imposes a predefined funnel-shaped restraint for ligand binding and unbinding,\cite{limongelli2013,raniolo2020} whereas SinkMeta enhances sampling through adaptive localized sink potentials.\cite{Pan2025SinkMeta} Unlike these approaches, the adaptive confinement tube does not prescribe the confinement geometry a priori. Instead, our method defines an adaptive neighborhood around the provided reference path according to a local free-energy criterion. It is also distinct from, and complementary to, adaptive-path and path-metadynamics methods, where the path itself is updated during sampling.\cite{leines2012,perezdealbaortiz2018} In the present formulation, the reference path remains fixed, while the tube width is adjusted thermodynamically.

We first lay the theoretical groundwork that underlies the proposed method, prior to detailing its practical implementation. Next, we examine five two-dimensional model potentials to disentangle the geometric and thermodynamic effects of the adaptive confinement. We then evaluate the approach on $N$-acetyl-$N^\prime$-methylalanylamide (NANMA) isomerization in vacuum, a common molecular benchmark, before applying it to two larger systems, including the folding of the mini-protein chignolin in water, and the folding-upon-binding of the ribose-binding protein (RBP). The method is implemented in the Colvars open-source library,\cite{fiorin2013,fiorin2024} and is, thus, compatible with multiple molecular-simulation engines, including NAMD, LAMMPS, GROMACS, and Tinker-HP.

\section{\label{Sec:Theory}Theory}

This section establishes the theoretical basis of the adaptive confinement tube. For a locally harmonic orthogonal well, we show that a tube placed around a fixed free-energy contour captures an $s$-independent fraction of the orthogonal partition function. In the hard-wall limit, the confinement, therefore, contributes only an additive constant to the path-projected free energy, and does not affect the free-energy differences along $s$.

Let $\{R_i\}_{i=1}^{M}$ denote an ordered set of equidistant reference configurations, and let $d(R,R_i)$ be a metric, typically a mean-squared displacement after optimal alignment. In the arithmetic PCV of Branduardi et al.,\cite{branduardi2007} the progress and the orthogonal coordinates are defined as,

\begin{equation}
\left\{
\begin{array}{lll}
s(R) &=& \displaystyle \frac{1}{M-1}\,
        \frac{\displaystyle \sum_{i=1}^{M} (i-1)\, e^{-\lambda d(R,R_i)}}
             {\displaystyle \sum_{i=1}^{M} e^{-\lambda d(R,R_i)}} ,
\\[2pt]
z(R) &=& \displaystyle  -\frac{1}{\lambda}
        \ln \sum_{i=1}^{M} e^{-\lambda d(R,R_i)} ,
\label{eq:pcv}
\end{array}
\right.
\end{equation}

where $\lambda$ controls the sharpness of the soft minimum. The coordinate $s$ measures progress along the reference path, whereas $z$ measures excursions from it. Because the metric is usually a squared distance, the arithmetic $z$ coordinate is not itself a physical radial displacement. This distinction matters for the volume-element correction discussed below.

Let $\rho(s,z)$ be the equilibrium density in the PCV coordinates and define the free energy,
\begin{equation}
F(s,z) = -\frac{1}{\beta} \ln \rho(s,z).
\end{equation}
with $\beta = 1/\kT$. The path-projected free energy corresponds to the marginal,
\begin{equation}
F(s) = -\frac{1}{\beta} \ln \int \dd z \; e^{-\beta F(s,z)},
\label{eq:Fs}
\end{equation}
A tube restraint restricts the allowed orthogonal region at each value of $s$. In the hard-wall limit, this gives the restricted marginal,
\begin{equation}
F_{\mathrm{tube}}(s)
  = -\frac{1}{\beta} \ln \int_{\Omega(s)} \dd z \; e^{-\beta F(s,z)} ,
\label{eq:Ftube}
\end{equation}
where $\Omega(s)$ is the orthogonal region of the tube. The corresponding captured fraction is given by,
\begin{equation}
p(s)
=
\frac{\displaystyle \int_{\Omega(s)} \dd z \; e^{-\beta F(s,z)}}
     {\displaystyle \int \dd z \; e^{-\beta F(s,z)}} .
\label{eq:pdef}
\end{equation}
Equations~\eqref{eq:Fs}--\eqref{eq:pdef} imply
\begin{equation}
F_{\mathrm{tube}}(s) = F(s) -\frac{1}{\beta} \ln p(s) .
\label{eq:bias}
\end{equation}
Thus, a tube with an $s$-independent captured fraction shifts the projected free energy by a constant, whereas an $s$-dependent captured fraction distorts free-energy differences and apparent barriers along the path.

To evaluate $p(s)$, we introduce local coordinates around the reference path. Let the path be embedded in an effective configuration or descriptor space of dimension $D_{\mathrm{eff}}$, hereafter referred to as effective perpendicular dimensionality. Locally, one direction is tangential to the path and parametrized by $s$, while the remaining
$d_\perp = D_{\mathrm{eff}} - 1$
directions are transverse to it. A configuration lying near the path can, therefore, be decomposed into a tangential displacement and a perpendicular vector $\boldsymbol{\eta}\in\mathbb{R}^{d_\perp}$, with a radial magnitude
$\varepsilon = \|\boldsymbol{\eta}\| \ge 0$. Under local isotropy, the angular
degrees of freedom can be integrated out, giving the radial volume element,
\begin{equation}
\dd \boldsymbol{\eta}
\propto
\varepsilon^{d_\perp-1}\,\dd\varepsilon .
\end{equation}

For a path in a two-dimensional $(\mathrm{CV_1}, \mathrm{CV_2})$ space,
$D_{\mathrm{eff}}=2$, and, hence, $d_\perp=1$. For a PCV defined by a
higher-dimensional descriptor, $D_{\mathrm{eff}}$ should be understood as the local effective rank of the fluctuations sampled around the path, rather than the formal number of Cartesian coordinates.

Near the path, we approximate the physical orthogonal free energy by a harmonic form,
\begin{equation}
\Delta V(\varepsilon,s)
=
\frac{1}{2}\kappa(s)\varepsilon^2 ,
\qquad
\kappa(s)>0 ,
\label{eq:harm}
\end{equation}
where $\kappa(s)$ is the local perpendicular stiffness. The unconfined orthogonal partition function then scales as $\kappa(s)^{-d_\perp/2}$, giving the physical perpendicular entropic contribution of
$
1/2 \times d_\perp \beta^{-1} \times \ln \kappa(s)
$
to the marginal free energy, up to an $s$-independent constant. This contribution is part of the correct path-projected free energy, and should not be removed by the tube construction.

For a tube of physical radius $\varepsilon_{\max}(s)$, the captured fraction is,
\begin{equation}
\begin{array}{lll}
p(s)
&=&
\frac{\displaystyle
\int_{0}^{\varepsilon_{\max}(s)}
\dd\varepsilon\,
\varepsilon^{d_\perp-1}
e^{-\frac{1}{2}\beta\kappa(s)\varepsilon^{2}}
}
{\displaystyle
\int_{0}^{\infty}
\dd\varepsilon\,
\varepsilon^{d_\perp-1}
e^{-\frac{1}{2}\beta\kappa(s)\varepsilon^{2}}
}
\\[0.8cm]
&=& \displaystyle
P\!\left(
\frac{d_\perp}{2},
\frac{1}{2}\beta\kappa(s)\varepsilon_{\max}^{2}(s)
\right),
\label{eq:p}
\end{array}
\end{equation}
where $P(a,x)=\gamma(a,x)/\Gamma(a)$ is the regularized lower incomplete gamma function.\cite{olver2010nist}

A fixed tube, $\varepsilon_{\max}(s)=\varepsilon_0$, does not in general preserve a constant captured fraction. The argument of $P$ in  \eqref{eq:p} is proportional to $\kappa(s)\varepsilon_0^2$, so soft and stiff regions retain different fractions of the orthogonal partition function. Consequently, the tube-induced contribution $-1/\beta \ln p(s)$ varies along the path and changes the apparent free-energy profile.

An adaptive confinement tube instead defines the wall by a fixed free-energy excess $\dF$ above the local minimum of the orthogonal well,
\begin{equation}
\frac{1}{2}\kappa(s)\varepsilon_{\max}^{2}(s) = \dF .
\label{eq:contour_condition}
\end{equation}
Equivalently,
\begin{equation}
\varepsilon_{\max}(s) = \sqrt{\frac{2\dF}{\kappa(s)}} .
\label{eq:contour}
\end{equation}
Substitution into  \eqref{eq:p} gives,
\begin{equation}
p
=
P\!\left(
\frac{d_\perp}{2},
\frac{1}{\beta} \dF
\right),
\end{equation}

\noindent from which it follows,

\begin{equation}
F_{\mathrm{tube}}(s)-F(s)
=
-\frac{1}{\beta}\ln p
=
\mathrm{const}.
\label{eq:invariance}
\end{equation}
The dependence on the local stiffness, $\kappa(s)$, cancels identically. Therefore, for a locally harmonic orthogonal well of constant effective perpendicular dimensionality, a fixed free-energy contour captures the same fraction of the orthogonal partition function at every point along the path. The tube width adapts geometrically, but the retained orthogonal configurational-space fraction is controlled thermodynamically by $\dF$.

This result clarifies the role of $\dF$. It is not a geometric distance cutoff, but the free-energy cost incurred by the largest orthogonal fluctuation within the tube. Equivalently, it sets the wall position in units of the local thermal width,
\begin{equation}
\frac{\varepsilon_{\max}(s)}{\sigma(s)}
=
\sqrt{2\beta\dF},
\qquad
\sigma(s) = [\beta\kappa(s)]^{-1/2},
\end{equation}
independently of the local stiffness. If the effective perpendicular dimensionality varies along the path,  \eqref{eq:invariance} acquires an additional $s$-dependence through $d_\perp(s)$, and the cancellation is no longer exact.

The derivation above is written in terms of the physical radial displacement, $\varepsilon$. In an arithmetic PCV, however, the sampled coordinate is approximately quadratic in this displacement, $z\propto\varepsilon^2$. The change of variables from $\varepsilon$ to $z$ introduces a metric correction that depends on the density,
\begin{equation}
\varepsilon^{d_\perp-1}\,\dd\varepsilon
\propto
z^{(d_\perp-2)/2}\,\dd z .
\label{eq:jacobian}
\end{equation}
Consequently, the conditional free energy obtained directly from a histogram in $z$ contains a coordinate-volume contribution. To target a physical free-energy contour in $\varepsilon$, rather than a contour distorted by the $z$ measure, this metric correction, in the form of a Jacobian term, must be accounted for. 
Details of the change-of-variables derivation and the corresponding correction for different PCV conventions are relegated to the supplementary material.


The contour invariance is exact under the aforementioned    assumptions, namely (i) a locally harmonic and locally isotropic orthogonal well, (ii) a hard-wall tube, and (iii) constant effective perpendicular dimensionality. In realistic systems, however, these assumptions may not necessarily hold. Anharmonicity can make the captured fraction weakly dependent on \(s\), finite-wall stiffness smoothes the hard-wall limit, and poorly sampled flat regions may thwart convergence of the empirical confinement potential. Moreover, the captured fraction should not be interpreted as the fraction of the unrestricted \(z\)-space retained by the tube, but rather as the fraction of the orthogonal basin corresponding to the path that remains meaningful for the chosen PCV. Accordingly, the denominator in \eqref{eq:pdef} is best understood as the ideal local partition function reflecting the reference path, which excludes off-path configurations for which \(s\) becomes ill-conditioned, or for which alternative channels are sampled. If the admissible perpendicular displacement \(\varepsilon\) is properly selected, however, the sampled region remains sufficiently local that the perpendicular free-energy surface can again be approximated as harmonic, isotropic, and channel-specific, thereby recovering the conditions under which the constant-fraction argument is valid. The practical implementation, therefore, estimates the confinement from the local \(z\) distribution rather than from the harmonic expression alone. The harmonic result should be understood as the limiting argument explaining why a free-energy-based adaptive tube, constrained to the locally valid path-associated domain, is less arbitrary than a fixed geometric cutoff.


\section{Method}
\label{sec:methods}
\label{sec:wall}\label{sec:engine}\label{sec:ramp}\label{sec:zfloor}\label{sec:recovery}

The adaptive confinement tube is implemented in Colvars as a geometric restraint defined on the PCVs $(s,z)$. This restraint does not bias the sampling along the path directly. Instead, it confines the orthogonal coordinate within an $s$-dependent tube, thereby defining the path-local ensemble, in which the free-energy profile is estimated.

Confinement is applied to the offset coordinate,
\begin{equation}
\delta z = z - z_{\mathrm{floor}}(s),
\end{equation}
where $z_{\mathrm{floor}}(s)$ is the on-path baseline of the PCV. This baseline is necessary because, for an arithmetic-path variable, the on-path value of $z$ is the soft-min floor rather than zero. 

\begin{figure}[t]
\centering
\includegraphics[width=\columnwidth]{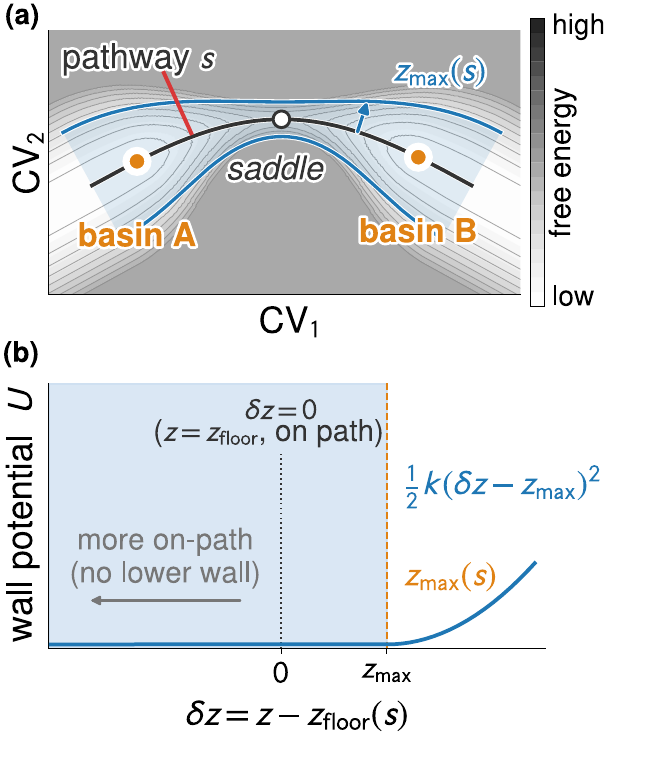}
\caption{\label{fig:schematic}Geometry of the adaptive confinement tube.
(a)~A reference path defines the progress coordinate, $s$, while the transverse coordinate, $z$, measures excursions from the path. The adaptive envelope, $z_{\max}(s)$, confines sampling to a path-local tube. Because the wall follows a fixed orthogonal free-energy contour, the tube widens in soft regions and narrows in stiff regions.
(b)~At fixed $s$, confinement is applied through a one-sided flat-bottom restraint in the offset coordinate $\delta z=z-z_{\mathrm{floor}}(s)$. The restraint is zero inside the tube, and harmonic outside it. No lower wall is imposed, because a smaller $\delta z$ corresponds to configurations closer to the reference path.}
\end{figure}

The tube is imposed through a one-sided flat-bottom restraint, (Fig.~\ref{fig:schematic}),
\begin{equation}
U(z,s) =
\begin{cases}
\displaystyle
\frac{1}{2}\,k\,\big(\delta z - z_{\max}(s)\big)^{2},
   & \delta z > z_{\max}(s),\\[3pt]
0, & \text{otherwise},
\end{cases}
\label{eq:wall}
\end{equation}
with force constant $k$. Only the outward side of the tube is restrained. configurations with smaller $\delta z$ are closer to the reference path, and, therefore, do not require a lower wall. The functions $z_{\mathrm{floor}}(s)$ and $z_{\max}(s)$ are represented on a grid in $s$, and smoothed to provide a continuous envelope with bounded gradients.

The same framework supports both fixed and adaptive confinements. In the fixed mode, $z_{\max}(s)$ is prescribed and held constant during the simulation, recovering the conventional fixed half-harmonic tube used as a baseline in this work. In the adaptive mode, $z_{\max}(s)$ evolves toward a prescribed free-energy contour of the local orthogonal distribution,
\begin{equation}
\widehat F(\delta z\mid s)-\widehat F(0\mid s)=\dF .
\label{eq:deltaFstar}
\end{equation}
The free-energy contour is estimated from the empirical conditional distribution of $\delta z$ in each bin of $s$, with the Jacobian correction described in \nameref{Sec:Theory} and in the supplementary material, when the sampled coordinate is a PCV. When local sampling is insufficient, the harmonic estimate,
\begin{equation}
z_{\max}(s) \simeq \sigma_{\delta z}(s)\sqrt{2 \beta \dF}
\end{equation}
is used as a fallback value. The update is regularized by smoothing along $s$ and by delaying histogram-based adaptation until sufficient local statistics has been accrued in any given bin. These safeguards allow the confinement potential to be initialized from a guess width, while avoiding noisy early updates.

This construction targets a free-energy-defined tube rather than a distance-defined one. For a locally harmonic orthogonal well, the contour condition in \eqref{eq:deltaFstar} is equivalent to retaining a constant fraction of the orthogonal partition function, as shown in \nameref{Sec:Theory}. The adaptive wall, therefore, changes its width along the path, all things being equal.

The tube restraint is independent of the enhanced-sampling method used to explore the progress coordinate. In this work, sampling along $s$ is driven primarily by well-tempered metadynamics (WT-MtD) and its combination with the extended-Lagrangian version of ABF, WTM-eABF.\cite{barducci2008,lesage2017,fu2016,fu2018} For the broad-collapsed model surface, we use WT-MtD alone, and for RBP, we turn to the on-the-fly probability enhanced sampling (OPES) method.\cite{invernizzi2020} In a nutshell, the adaptive tube controls the orthogonal extent of the sampled ensemble, whereas the chosen enhanced-sampling bias accelerates motion along the path.

For WTM-eABF simulations, the path-projected free energy is estimated using the corrected $z$-averaged restraint (CZAR) estimator on the gradients.\cite{lesage2017} For WT-MtD and OPES simulations, the profile is obtained by reweighting the time-dependent bias, using the Tiwary-Parrinello estimator for WT-MtD\cite{tiwary2015,bonomi2009} and the corresponding OPES weights, respectively.\cite{invernizzi2020} The resulting quantity is the tube-restricted marginal, $F_{\mathrm{tube}}(s)$. Because the confinement is time-dependent during the initial adaptation stage, free-energy estimates are computed from the post-adaptation portion of the trajectory.

The complete set of simulation parameters, tube-update settings, and implementation details are provided in the supplementary material.


\section{Results}
\label{sec:results}

To probe our method, we first tackle several two-dimensional model potentials, which isolate the mechanism and test the theory directly. We subsequently turn to more realistic molecular systems.
NANMA isomerization represents a common reference for a backbone-dihedral PCV, and chignolin folding and RBP folding-upon-binding processes highlight larger, real-life applications.

\subsection{Model potentials}
\label{sec:models}

Each model is a two-dimensional potential sampled by Langevin dynamics, with
WTM-eABF driving the progress coordinate of an arithmetic PCV. Runs are
$5\times10^{6}$-step long unless noted. Four of the potentials  target a single
facet of the multifaceted proposed approach. A fifth potential  depicts one surface, upon which no
fixed tube width can succeed, and is sampled by WT-MtD along $s$
alone. The default confinement target is set to be $\dF = 1\,\kT$, unless otherwise noted.

\subsubsection{M\"uller--Brown and triple-well potentials}

The M\"uller--Brown and triple-well surfaces both have roughly harmonic
perpendicular wells along the minimum-energy path, the regime in which the
contour invariance of \nameref{Sec:Theory} should hold. They let us overlay the
adapted wall on the surface, and confirm that the recovered $F(s)$ matches the analytical ground truth. They also test the invariance directly, through the captured
fraction as a function of the local stiffness, $\kappa(s)$, which a confinement tube
holds flat across all values accessible, where a fixed tube does not (\textcolor{blue}{Figure~S1}). The confinement adaptation mechanism recovers the analytical barrier within about $\pm0.4$~kcal/mol on both
surfaces (\tabref{tab:recovery}).


\begin{table}[t]
\caption{\label{tab:recovery}CZAR-recovered barrier of the contour adaptation mechanism against
the analytic truth (kcal/mol). All runs are $5\times10^{6}$ steps.}
\begin{ruledtabular}
\begin{tabular}{lccc}
potential & true barrier & steps & error \\
\hline
triple-well     & $2.6$ & $5\times10^{6}$ & $+0.1$ \\
M\"uller--Brown & $4.0$ & $5\times10^{6}$ & $-0.4$ \\
dual-channel    & $3.8$ & $5\times10^{6}$ & $+0.1$ \\
lake            & $1.5$ & $5\times10^{6}$ & $+0.3$ \\
\end{tabular}
\end{ruledtabular}
\end{table}

\subsubsection{Dual-channel potential and inter-channel leakage}

The dual-channel potential has two nearly degenerate channels that share the same
range of $s$ and pinch together at intermediate $s$, where a modest barrier in
the transverse coordinate $y$ separates them, forming an entropic hourglass. Geometric restraints too loose to confine sampling allow
it to hop through the bottleneck and to populate both channels. Table~S3 reports the fraction of frames in the wrong (lower) channel at central $s$ as the radius is varied. The loosest tube
($z_{\max}=0.8$) admits $38\%$ leakage. Tightening it suppresses the leakage monotonically with lower values of $z$. The default adaptive confinement tube reaches $0.3\%$ leakage, matching the
tightest fixed tube with no predefined width (\figref{fig:models}a,b).

\subsubsection{Lake potential and over-confinement distortion}

\begin{figure*}[t]
\centering
\includegraphics[width=\textwidth]{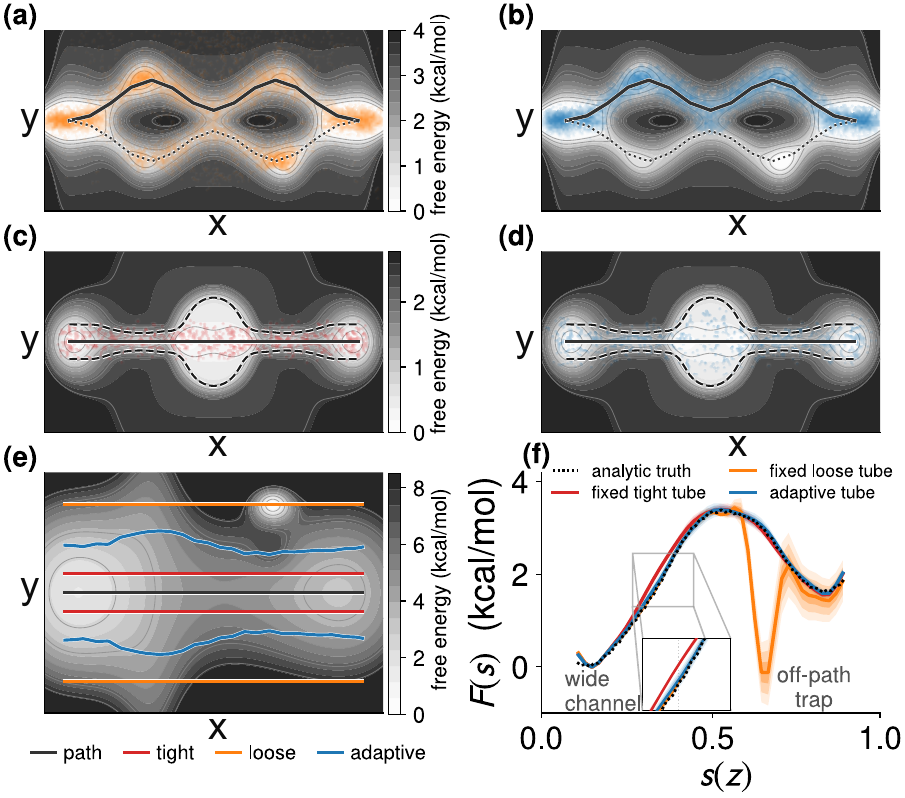}
\caption{\label{fig:models}The adaptive tube's response on three model surfaces.
(a)~a loose tube ($z_{\max}=0.8$) hops through the pinch, populating both channels.
(b)~the adaptive tube holds the upper channel, matching a tight fixed-$z$ tube.
(c)~a fixed-tight tube ($z_{\max}=0.02$) clips the wide basin, well inside the natural $\dF$-contour width (dashed).
(d)~the adaptive tube fills the lake and narrows at the necks, tracking the $\sim$25-fold width variation.
(e)~the BC potential $V(x,y)$ (gray) with a soft channel near $s\approx0.34$ and an off-path trap near $s\approx0.66$, overlaid with fixed tight (red), fixed loose (orange), and adaptive (blue) tube walls.
(f)~Path-projected $F(s)$ per tube (mean $\pm1$~SD, three seeds) vs.\ the trap-excluded reference (dotted): fixed tight over-confines, fixed loose leaks into the trap, adaptive tracks the truth throughout.}
\end{figure*}

The lake potential is a wide, flat basin connected to the rest of the path through narrow corridors, such that the natural perpendicular width varies by a factor of about
$25$ along $s$. It is the cleanest demonstration of the over-confinement
distortion described in \nameref{Sec:Intro}, one that can be shown with no
restraint at all. Clipping the perpendicular coordinate to a tight window
inverts the free-energy ordering of the lake and the basin, turning a real
entropic preference for the lake ($-0.4$~kcal/mol) into an apparent
preference for the basin ($+0.1$~kcal/mol), an altogether swing of $0.4$~kcal/mol
(Table~S4, \figref{fig:models}c,d). A loose window preserves the ordering, and the adaptive confinement tube recovers the correct $F(s)$ hierarchy (\tabref{tab:recovery}).

\subsubsection{Broad-collapsed potential, a single surface where no fixed width works}

The dual-channel and lake potentials discriminate the two failure modes one at a time.
The BC potential combines them in a single surface, so that no single
fixed width can avoid both. Its on-path channel is wide and soft near
$s\approx0.34$, and narrow and stiff elsewhere. A deep off-path trap sits
behind a low-guard wall near $s\approx0.66$ (\figref{fig:models}e). A tube tight enough to exclude the trap clips the wide channel, and a tube wide enough to fill
the channel encompasses the trap. Under exact projection of the landscape, free of sampling error, a fixed-tight tube over-confines the wide channel, and increases
$F(s)$ by about $0.3$~kcal/mol, while a fixed-loose tube allows sampling to leak into the trap
and causes $F(s)$ to decrease sharply near $s\approx0.66$ by roughly $3$~kcal/mol. The adaptive
tube opens in the channel and pinches near the trap, staying within about
$0.1$~kcal/mol of the trap-excluded truth at both loci. The recovered free-energy profiles confirm
this observation across three independent metadynamics seeds (\figref{fig:models}f). The
fixed-tight profile rides above the truth on the rising flank into the channel,
the fixed-loose profile collapses at the trap, and only the adaptive profile
tracks the truth throughout.

\subsection{NANMA isomerization}
\label{sec:nanma}

\begin{figure}[t]
\centering
\includegraphics[width=\columnwidth]{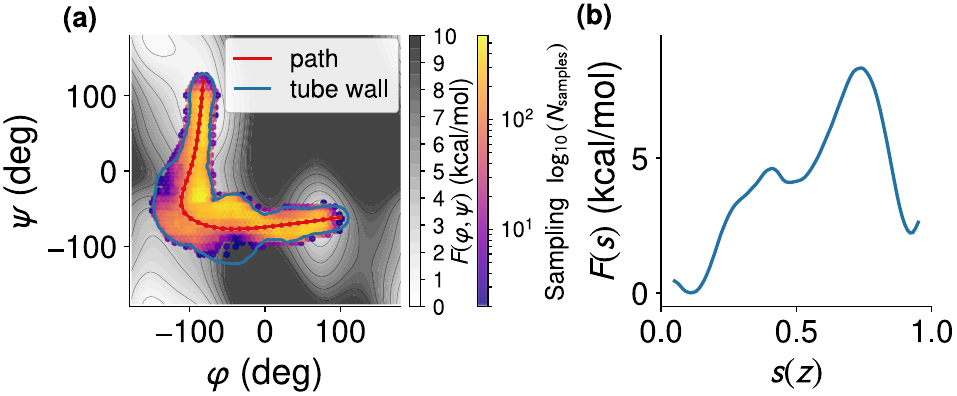}
\caption{\label{fig:nanma}NANMA isomerization explored in $(\varphi,\psi)$ dihedral space. (a)~Ramachandran free-energy surface $F(\varphi,\psi)$ (gray) overlaid
with the sampled density (color, $\log_{10}N_{\mathrm{samples}}$), the reference path (red), and
the adapted tube wall (blue) of the $\dF=3\,\kT$ run. (b)~Path-projected free
energy $F(s)$ from the adaptive tube (blue). The tube wall in (a) embraces the path with a width
that broadens over its interior and contracts towards the endpoints.}
\end{figure}

NANMA isomerization in vacuum is a very common molecular benchmark for enhanced-sampling methods. We
optimized a pathway in the $(\varphi,\psi)$ dihedral space, along which PCV sampling was performed using
WTM-eABF. The adaptive confinement tube reproduces the well-known reference $F(s)$ (\figref{fig:nanma}), returning a transition free-energy barrier of 
$\Delta F \approx8.3$~kcal/mol and a $\Delta\Delta{F} \approx 2$ kcal/mol between the so-called C$_{\rm 7ax}$ and C$_{\rm 7eq}$ conformational states, in agreement with previous studies.\cite{branduardi2007,Tang2025SMwSTTutorial}

\subsection{Chignolin fast-folding}
\label{sec:chignolin}

\begin{figure}[t]
\centering
\includegraphics[width=\columnwidth]{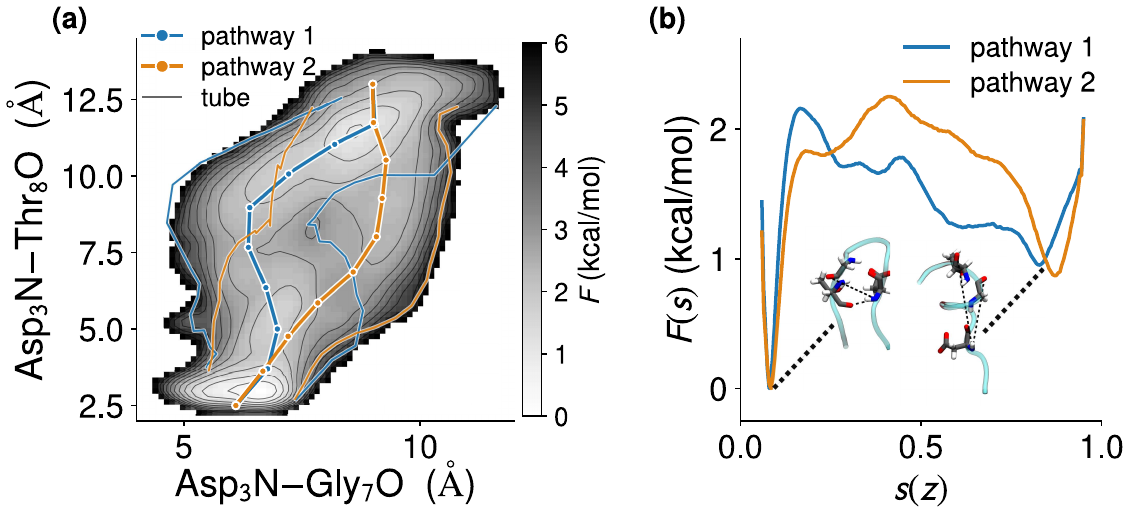}
\caption{\label{fig:chignolin}Chignolin reversible folding with the adaptive tube along two alternate pathways. (a) Two-dimensional free-energy landscape projected in the plane formed by the
backbone hydrogen-bond distances Asp$_3$N--Gly$_7$O and Asp$_3$N--Thr$_8$O, with the two paths and their projected
tube walls overlaid on the grayscale surface. (b) Folding free-energy profile, $F(s)$, along each pathway, with the folded-state basin anchored at zero.}
\end{figure}

Chignolin is a hydrated mini-protein, and its reversible folding has been the subject of  extensive computational investigation.\cite{maier2015,phillips2020} It is a realistic case, wherein a fixed
$\zmax$ is hard to guess beforehand, and the ill-conditioning failure mode is real, making it a natural benchmark for the present method. The folding pathways are supplied by the path-committor-consistent artificial neural network approach,\cite{megias2025}
in the CV subspace formed by the Asp$_3$N--Gly$_7$O and Asp$_3$N--Thr$_8$O distances, which are
key backbone hydrogen bonds.\cite{Oshima2019replica,Chen2021,kang2024Parrinello,megias2025}
\figref{fig:chignolin} shows the folding free-energy profile, $F(s)$, together with the free-energy landscape projected in the plane of the two hydrogen bonds. The free-energy differences between the end states, and the corresponding transition barrier height, agree with reference values from the literature---the folded-state  basin being around $1$~kcal/mol more stable than the unfolded one.\cite{harada2011,Oshima2019replica,Chen2021,yang2024learning,megias2025}

\subsection{Ribose-binding protein folding-upon-binding}
\label{sec:rbp}

\begin{figure}[t]
\centering
\includegraphics[width=\columnwidth]{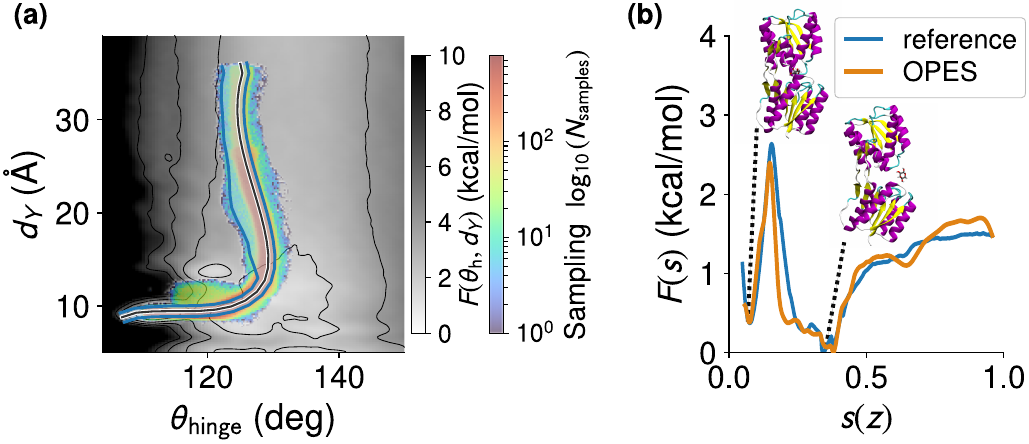}
\caption{\label{fig:rbp_opes}Folding-upon-binding in RBP sampled with OPES along the confinement tube. (a) Two-dimensional
OPES free energy in the plane of the hinge-bending angle $\theta_{\mathrm{hinge}}$ and the
ligand displacement $d_Y$ (grayscale surface with 2~kcal/mol contours), overlaid with
the combined OPES sampling density, $\log_{10}(N_{\mathrm{samples}})$, the path
center line and its projected adaptive tube walls, which form a funnel that is
narrow at the bound state, and widens toward the unbound state. (b) One-dimensional
free-energy profile, $F(s)$, determined along the pathway: The marginalized two-dimensional OPES reference
and OPES PCV sampling.}
\end{figure}

RBP is a periplasmic receptor, which captures ribose
for bacterial transport and chemotaxis.\cite{bjorkman_probing_1994} Its ligand reversible binding is coupled to an open-to-closed hinge-bending conformational rearrangement of its two domains,
a transition characterized in earlier computational studies.\cite{ren_unraveling_2021}
The reference path is the minimum-free-energy path through the two-dimensional
surface of Kang et al.,\cite{kang2026convergence} spanned by the
interdomain hinge angle, $\theta$, and the distance, $r$, between the ribose and the binding site,
connecting the holo-closed and apo-open states through a holo-open intermediate.
Binding is the regime whereby the fixed-width dilemma is the most acute. The natural
perpendicular width is large in the unbound state, and narrow in the bound state,
so that a single fixed tube either over-confines the unbound-state ensemble, or, alternatively, under-confines the bound-state ensemble, representing a good test case for a funnel-like $z$ constraint (see \figref{fig:rbp_opes}a). The orthogonal tube keeps
the system near this path, while $F(s)$ is computed along it, and the projected free-energy profile is compared against the free-energy surface marginalized over the perpendicular
coordinate, $z$. We have opted for OPES-based sampling to showcase the adaptability of the method with a different enhanced-sampling scheme. The pathway was obtained along the minimum free-energy path at zero temperature from the two-dimensional free-energy surface of Kang et al.,\cite{kang2026convergence}. We observe that $F(s)$ computed from the adaptive confinement approach, shown in \figref{fig:rbp_opes}b, matches the reference values obtained from the marginalization of the two-dimensional reference free-energy landscape along the same pathway.

\section{Limitations}
\label{sec}

The scope of the confinement-tube construction is determined by the assumptions underlying the theory and by the quality of the reference path. First, the method assumes that the reference path provides a meaningful local description of the transition channel. Because $z$ measures the distance from the path rather than the direction of the excursions, an ill-constructed reference path may still ultimately relax towards a lower-free-energy route. In such cases, tube adaptation improves the local definition of the sampled ensemble, but is not a substitute for a physically relevant path.

Second, confinement adaptation requires sufficient exploration of the local orthogonal distribution. In flat or weakly confined regions, where the perpendicular stiffness approaches zero, a narrowly initialized tube may not extend far enough to identify the target contour. These cases require a broader initial confinement, longer local sampling, or additional diagnostics of the empirical $z$ distribution.

Third, the adaptive confinement is not, per se, a general criterion for separating competing transition channels. If two channels have similar free energy and are not separated by a clear barrier along $z$, then the sampled configurations cannot be assigned unequivocally to one specific mechanism.

Finally, the contour-invariance result (see \nameref{Sec:Theory}) is exact only in the hard-wall limit for a locally harmonic, locally isotropic orthogonal well, with constant $D_{\rm eff}$. Anharmonicity, anisotropy, finite-wall stiffness, variation of $d_\perp$ along the path, and incomplete sampling can introduce deviations from the ideal cancellation. The method, therefore, reduces, but does not eliminate, the need to check convergence of both the adaptive confinement and the free-energy profile along $s$.

\section{Discussion and conclusions}
\label{sec:discussion}

The present work introduces an adaptive confinement tube for PCV sampling.\cite{branduardi2007} The central idea is to define orthogonal confinement by a local free-energy criterion rather than by a fixed geometric distance. The theoretical result in \nameref{Sec:Theory} explains why this distinction matters. For a locally harmonic, orthogonal well, a tube placed at a fixed free-energy contour captures an $s$-independent fraction of the orthogonal partition function. In the hard-wall limit, the confinement, therefore, contributes only an additive constant to the path-projected free energy, leaving free-energy differences along the path unchanged. A fixed-width tube has no such invariance, because the retained fraction of orthogonal configurational space changes with the local perpendicular stiffness.

The practical consequence is that the method replaces a fragile geometric parameter, namely the tube width, with a physically interpretable energy scale, $\dF$. The tube can widen in soft basins and narrow near stiff-transition regions without imposing an arbitrary global distance cutoff. This approach reduces two common failure modes of PCV sampling: (i) Geometric instability when the trajectory wanders too far from the reference path, and (ii) thermodynamic distortion when a fixed tube over-confines natural perpendicular fluctuations.

The underlying theory of our method has broader implications than the specific PCV implementation considered here. Other geometric restrictions that remove orthogonal configurational space, such as funnels, cylinders, or shaped pockets, could in principle be formulated in terms of free-energy contours rather than prescribed boundaries. Such extensions would require separate validation, especially when the orthogonal space is anisotropic, or when multiple channels are not well represented by a single radial coordinate. Within the present setting, however, the adaptive confinement tube provides a natural, thermodynamically grounded alternative to fixed-width confinement.

An obvious extension of the present work would consist in combining confinement-tube adaptation with adaptive-path\cite{perezdealbaortiz2018} or path-metadynamics\cite{leines2012} schemes. In such a framework, the reference path would evolve to improve the representation of the transition mechanism, while the tube width would adapt to maintain a controlled orthogonal ensemble around the current path. This combination could provide a route toward simultaneous pathway optimization and thermodynamically consistent confinement, while preserving the separation between progress along the path and excursions away from it.


\section*{Supplementary material}
See the supplementary material for the implementation and reproducibility
details, the convergence diagnostics, and the sensitivity scans in the confinement parameters.

The Colvars code containing the adaptive $z$ restraint implementation can be found at
\url{https://github.com/Lapsis-glitch/colvars}.

\begin{acknowledgments}
C.C. acknowledges the European Research Council (project 101097272 ``MilliInMicro''). C.C. also acknowledges the Agence Nationale de la Recherche under France 2030 (contract ANR-22-PEBB-0009) for support in the context of the MAMABIO project (B-BEST PEPR). The authors of reference \citenum{kang2026convergence} are gratefully acknowledged for provision of the reference two-dimensional free-energy profile data for the ribose-binding protein folding upon binding.
\end{acknowledgments}

\section*{Author Declarations}
\subsection*{Conflict of Interest}
The authors have no conflicts to disclose.

\subsection*{Author Contributions}
\textbf{Radu A. Talmazan}, Conceptualization, Methodology,
Software, Writing.
\textbf{Cheng Giuseppe Chen}, Investigation, Methodology,
Writing.
\textbf{Chenyu Tang}, Investigation, Methodology,
Writing.
\textbf{Christophe Chipot}, Conceptualization, Funding acquisition, Supervision, Writing -- review and editing.

\section*{Data Availability Statement}
The data that support the findings of this study are available within the
article and its supplementary material. The adaptive tube restraint is
implemented in the open-source Colvars library.\cite{fiorin2013,fiorin2024} The
two-dimensional sandbox scripts and the input files for the model-potential,
alanine-dipeptide, chignolin, and RBP calculations are deposited in a public
repository on GitHub (
\url{https://github.com/Lapsis-glitch/colvars}).

\bibliography{refs}

\end{document}